# Tailoring Charge-Transfer at Metal-Organic Interfaces Using Designer Shockley Surface States

Anubhab Chakraborty,[1] Oliver L.A. Monti[1,2]*

[1]Department of Chemistry and Biochemistry, University of Arizona, Tucson, Arizona 85721, United States; email: anubhabc@arizona.edu

[2]Department of Physics, University of Arizona, Tucson, Arizona 85721, United States; email: monti@arizona.edu

Abstract

Metal-organic interfaces determine critical processes in organic electronic devices. The frontier molecular orbitals (highest occupied and lowest unoccupied molecular orbital, HOMO and LUMO) are crucial in determining charge-injection and -collection processes into and from the organic semiconductor films. Here we show that we are able to tune the interfacial electronic structure of a strongly interacting interfacial system formed by adsorption of the electron acceptor 1,4,5,8,9,11-hexaazatriphenylenehexacarbonitrile (HATCN, $C_{18}N_{12}$) on Ag thin films on Cu(111). The thickness-dependent Shockley surface state emerging on this layered metallic system couples to the LUMO, which allows precise control over the energetic position and filling of the charge-transfer interface state relative to the Fermi level ($E_F$). Our ability to tune the interfacial electronic structure while maintaining the structure of the molecular film represents an important step towards designing organic semiconductor interfaces.

Interfaces formed by π-conjugated organic molecules in contact with metal substrates have been extensively explored as building blocks to create a wide variety of organic electronic devices.[1–7] Among the diverse physics of the formation of an interface between these two materials classes, energy level alignment and hybridization of molecular and metal states are essential for determining the efficiency of charge carrier injection or extraction at the interface and play a key role in device performance.[8,9] Since one of the key actors for charge-transfer is the energetic difference between the metal Fermi level ($E_F$) and the frontier orbitals (HOMO, LUMO) of the organic molecule layer, it is desirable to create interfacial systems with tunable states that can allow control over interfacial charge-transfer and the associated barrier. One possibility is to choose an appropriate organic semiconductor to maximize orbital overlap and charge-transfer to or from the metal substrate. Equally important is the density of states (DOS) of the metal substrate near $E_F$, which can affect the interfacial charge-transfer and molecular self-assembly, as shown previously.[10,11]

In this work, we use the latter approach. We present an electronic structure investigation of the metal-organic interface formed by 1 monolayer (ML, ≈ 5.6 × $10^{13}$ molecules/cm$^2$)[12] of the organic semiconductor 1,4,5,8,9,11-hexaazatriphenylenehexacarbonitrile (HATCN) on thin films of Ag on a Cu(111) substrate. The HATCN molecule is a known electron acceptor on coinage metal surfaces,[13–15] and the energetic proximity of the HATCN LUMO to the Ag(111) $E_F$ allows interfacial charge-transfer. We combine this molecular system with the Ag/Cu(111) heterostructure, which is a well-studied metal thin film system with thickness dependent DOS near $E_F$. This allows precise tailoring of interfacial interactions. Our Ultraviolet Photoemission Spectroscopy (UPS) and Angle-Resolved Photoemission Spectroscopy (ARPES) results show the formation of an interfacial charge-transfer state in this system, and we demonstrate for the first time that the binding energy of this state can be tailored by controlling the Ag film thickness. Our method therefore provides a practical way of creating tunable charge-transfer interfaces which may enable tailoring of charge extraction or injection in organic electronics.

HATCN was commercially obtained (Alfa Chemistry, 99%) and purified by three cycles of gradient sublimation (553 K) in a custom-built vacuum furnace ($5 \times 10^{-6}$ Torr). The Cu(111) crystal was cleaned using repeated cycles of $Ar^+$ sputtering (1 keV, $5 - 10$ µA $cm^{-2}$) and annealing (800 K). Ag was deposited onto Cu(111) in a UHV sample preparation chamber ($7 \times 10^{-10}$ Torr) using a custom-built water cooled Knudsen source, and the deposition rate (~0.2 – 1 ML $min^{-1}$) was monitored using a quartz crystal microbalance (QCM). HATCN was also deposited using a custom-built water cooled Knudsen source at a rate of 0.2 ML $min^{-1}$. Our custom-built deposition system allows HATCN deposition at low temperatures (~375 K), well below the threshold required for $C_2N_2$ fragment desorption from HATCN.[15] Indeed, we do not observe defect formation in the adsorbed layer. All UPS, ARPES and Low-Energy Electron Diffraction (LEED) measurements were performed under UHV conditions ($2 \times 10^{-10}$ Torr) and at 298 K. The UPS and ARPES measurements were taken in a VG EscaLab MK II photoelectron spectrometer using a He I photon source (SPECS 10/35, $h\nu = 21.22$ eV), with an analyzer acceptance angle of $\pm 1.5°$ and a sample bias of -5 V. LEED images were acquired using an Omicron SPECTALEED instrument, and image analysis and distortion correction were done using LEEDCal and LEEDLab.[16]

We start by discussing the surface electronic structure of Ag thin films on Cu(111). This system has been extensively studied,[17–24] and the major spectral feature of interest for this work is the Shockley Surface State (SSS)[25] which appears as a result of the boundary conditions imposed by termination of the periodic crystal lattice at the surface. In coinage metals such as Cu(111) and Ag(111), the SSS exists in the projected band gap in the $\Gamma \rightarrow L$ direction of the band structure[26] and shows a free-electron-like dispersion. For the Ag/Cu(111) thin film system, the SSS evolves as a function of Ag thickness on the Cu(111) surface, as reported previously.[19,20,27] For a bare Cu(111) surface, the SSS has a binding energy of -0.32(1) eV (Figure 1(a), bottom panel). For sub-monolayer coverages up to one ML of Ag, two surface states are observed in the photoemission spectra, corresponding to the Cu(111) SSS and the newly emerging Ag SSS.[19,27,28] For Ag coverages ≥1 ML, the Cu(111) SSS disappears completely and only the Ag SSS feature is observed in the photoemission spectra. Importantly, the binding energy of the Ag SSS in this system depends on the thickness of the Ag film, and is most sensitive to the film thickness between 0-5 ML of Ag.[19,20,29] This is a result of the SSS wave function extending a few layers into the bulk and interacting with the Ag/Cu(111) interfacial potential.[27] Figure 1(a) and 1(b) show this thickness dependent binding energy behavior of the Ag/Cu(111) SSS. The binding energy shift is proportional to the amplitude-squared $|\Psi|^2$ of the SSS wavefunction at the Ag/Cu(111) interface and consequently follows an exponential decay:[30]

$$E_{SS}(d) = E_{SS}(\infty) + \left(E_{SS}(0) - E_{SS}(\infty)\right)e^{-2\beta d} \tag{1}$$

where $E_{SS}(d)$ is the SSS binding energy at an Ag film thickness of $d$ (measured in units of monolayer), $E_{SS}(\infty)$ and $E_{SS}(0)$ are the SSS binding energies of the Ag(111) and the Cu(111) surfaces respectively and $\beta^{-1}$ is the decay length of the SSS wavefunction into the bulk. Fitting our data to this model (Figure 1(b)) we get a decay length of 3.4(7) ML, which is consistent with literature.[19,20] The feature of importance here is that the SSS binding energy can be tuned between -0.32(1) eV (for 0 ML Ag, or Cu(111)) to -0.023(1) eV (for 35 ML Ag) by controlling the thickness of the Ag film, which provides an easy way to control the density of states (DOS) near $E_F$. Table 1 lists the binding energies of the $n$-ML Ag/Cu(111) for $n$ ranging from 0 to 35. The band structures of the $n$-ML Ag/Cu(111) measured using ARPES are shown in Figure S1(Supplementary Information). Notably, the half-width at half maximum (HWHM) of $n$-ML Ag/Cu(111) SSS (see Table 1) is larger for $n < 10$ and decreases with increasing Ag thickness, likely due to the localization of the SSS in the Ag layer for thicker films.[27] This reduces the broadening induced by electron scattering at the Ag/Cu(111) interface.

Next, we discuss the results of 1 ML of HATCN deposited on the $n$-ML Ag/Cu(111) system. HATCN is a strong electron acceptor, and previous studies[12,13,15,31,32] have

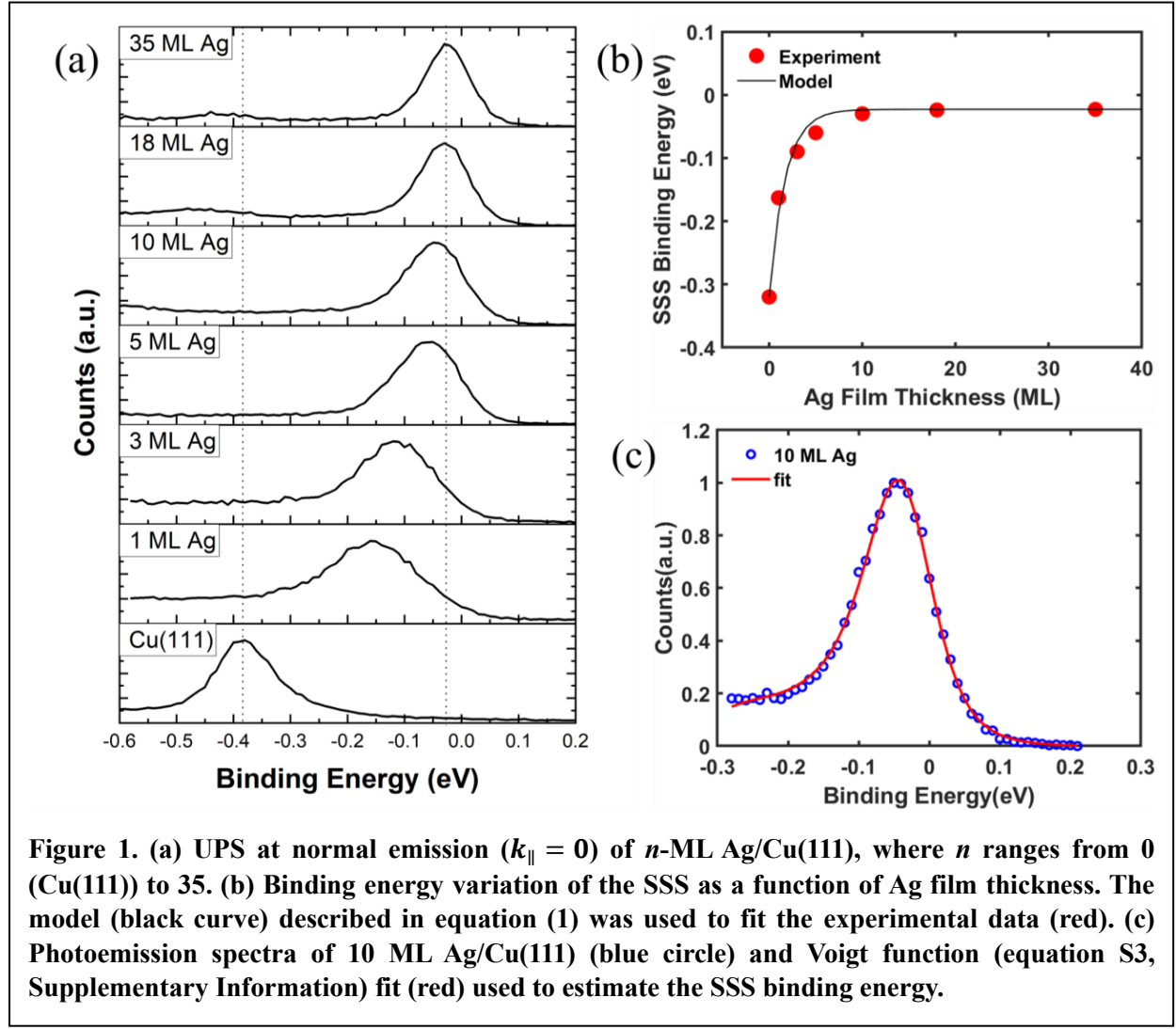

Figure 1. (a) UPS at normal emission ($k_\parallel = 0$) of $n$-ML Ag/Cu(111), where $n$ ranges from 0 (Cu(111)) to 35. (b) Binding energy variation of the SSS as a function of Ag film thickness. The model (black curve) described in equation (1) was used to fit the experimental data (red). (c) Photoemission spectra of 10 ML Ag/Cu(111) (blue circle) and Voigt function (equation S3, Supplementary Information) fit (red) used to estimate the SSS binding energy.

described the self-assembly and interfacial electronic properties of HATCN on the Ag(111) surface. At 1 ML coverage, the HATCN molecules adsorb with a face-on orientation on Ag(111) and form a well-ordered layer exhibiting a honeycomb structure[12,13] that corresponds to a (7 × 7) superstructure. Since the Ag layer on Cu(111) adopts largely the Ag(111) structure[18] with the natural lattice constant of Ag ($a_{Ag}$ = 0.289 nm) already at 1 ML, we expect the adsorption properties of HATCN / $n$-ML Ag/Cu(111) to be similar to HATCN/Ag(111). Indeed, our LEED results

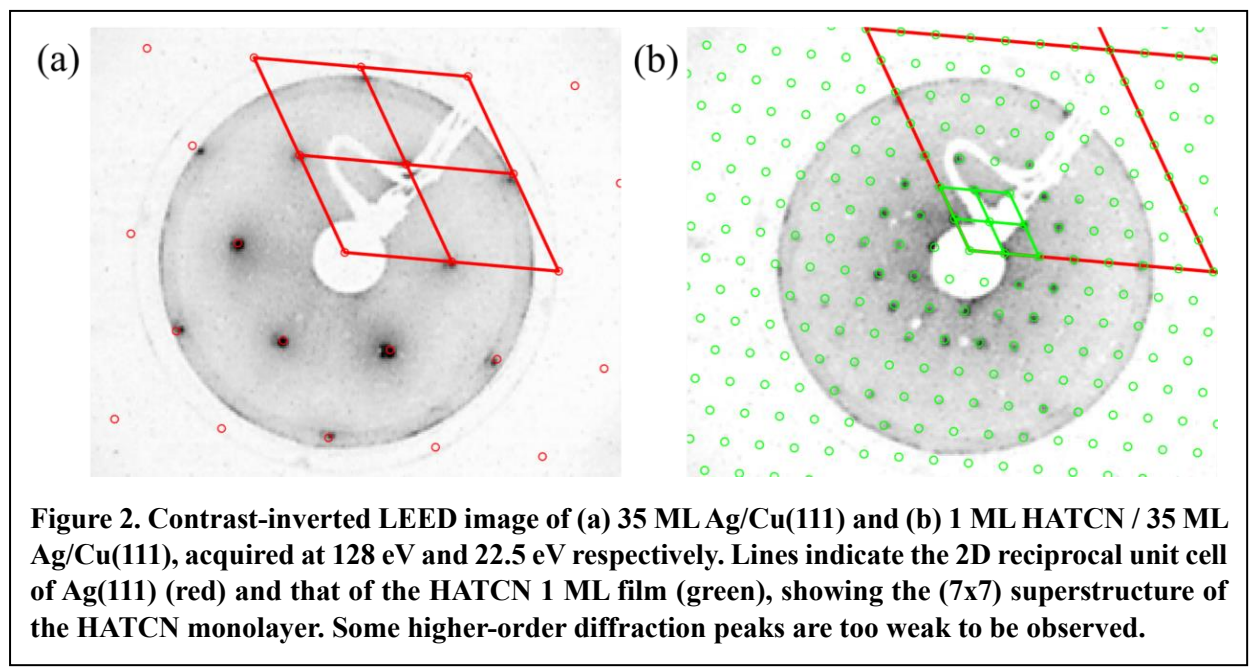

**Figure 2.** Contrast-inverted LEED image of (a) 35 ML Ag/Cu(111) and (b) 1 ML HATCN / 35 ML Ag/Cu(111), acquired at 128 eV and 22.5 eV respectively. Lines indicate the 2D reciprocal unit cell of Ag(111) (red) and that of the HATCN 1 ML film (green), showing the (7x7) superstructure of the HATCN monolayer. Some higher-order diffraction peaks are too weak to be observed.

(Figure 2(b)) of 1 ML HATCN / 35 ML Ag/Cu(111) show the expected (7 × 7) HATCN superstructure.

**Table 1. SSS Binding Energy and HWHM as a function of the Ag film thickness**

| Ag Film Thickness (ML) | SSS Binding Energy (eV) | SSS HWHM (eV) |
|---|---|---|
| 0 | -0.32(1) | 0.035(9) |
| 1 | -0.163(1) | 0.06(2) |
| 3 | -0.090(1) | 0.05(4) |
| 5 | -0.060(2) | 0.048(8) |
| 10 | -0.030(1) | 0.034(9) |
| 18 | -0.024(1) | 0.031(3) |
| 35 | -0.023(1) | 0.028(3) |

Due to the strong electron-accepting nature of HATCN, it chemisorbs on the Ag(111) surface and undergoes a charge-transfer of 1 electron from Ag into the HATCN LUMO to create a singly-occupied charge-transfer state, as reported

previously.[12,13,32] We shall refer to this state as the Hybrid Interface State (HIS)

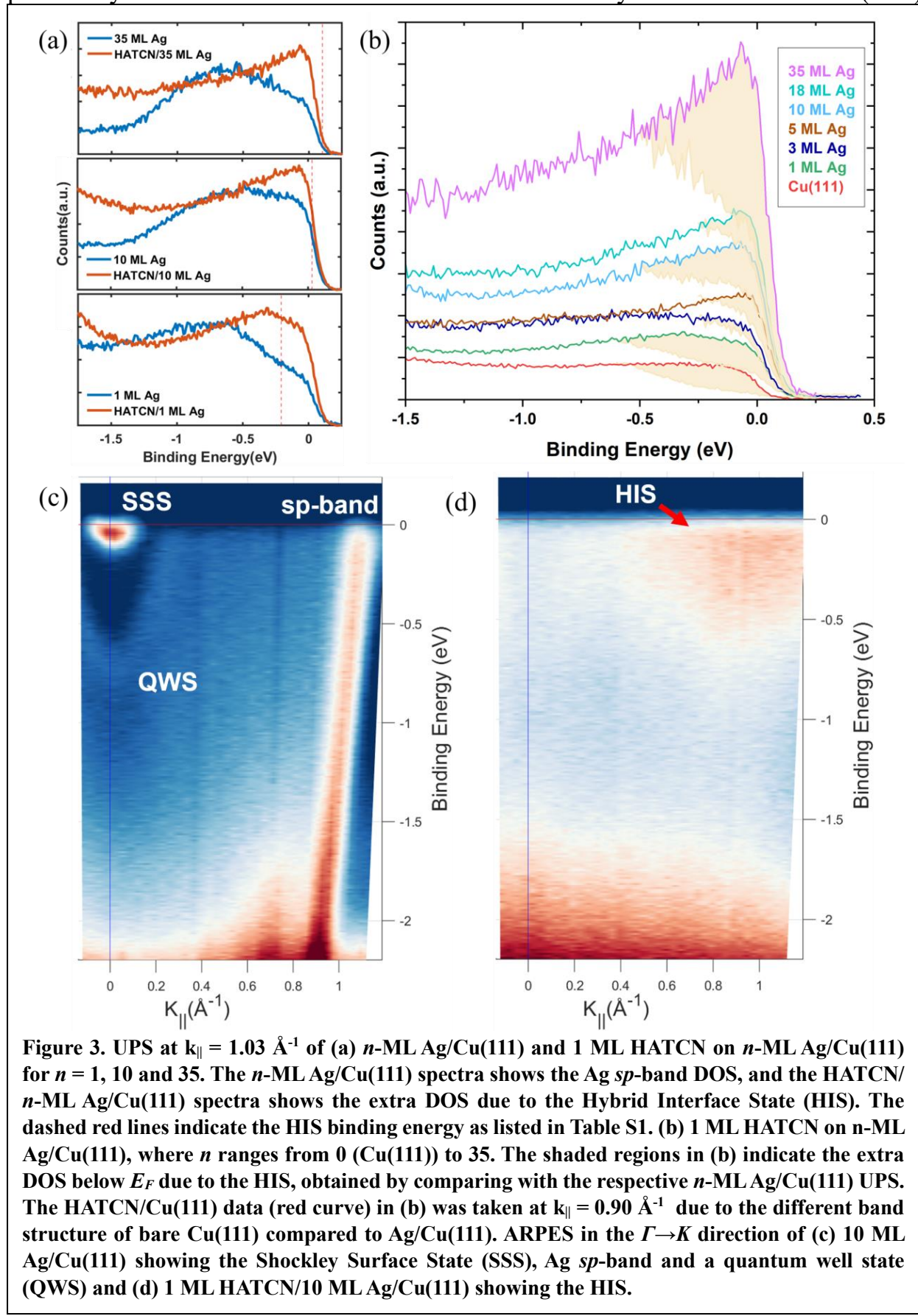

**Figure 3.** UPS at $k_{\parallel}$ = 1.03 Å$^{-1}$ of (a) *n*-ML Ag/Cu(111) and 1 ML HATCN on *n*-ML Ag/Cu(111) for *n* = 1, 10 and 35. The *n*-ML Ag/Cu(111) spectra shows the Ag *sp*-band DOS, and the HATCN/ *n*-ML Ag/Cu(111) spectra shows the extra DOS due to the Hybrid Interface State (HIS). The dashed red lines indicate the HIS binding energy as listed in Table S1. (b) 1 ML HATCN on n-ML Ag/Cu(111), where *n* ranges from 0 (Cu(111)) to 35. The shaded regions in (b) indicate the extra DOS below $E_F$ due to the HIS, obtained by comparing with the respective *n*-ML Ag/Cu(111) UPS. The HATCN/Cu(111) data (red curve) in (b) was taken at $k_{\parallel}$ = 0.90 Å$^{-1}$ due to the different band structure of bare Cu(111) compared to Ag/Cu(111). ARPES in the *Γ→K* direction of (c) 10 ML Ag/Cu(111) showing the Shockley Surface State (SSS), Ag *sp*-band and a quantum well state (QWS) and (d) 1 ML HATCN/10 ML Ag/Cu(111) showing the HIS.

henceforth. The term "former-LUMO (F-LUMO)" has also been used to refer to such states.[11,33] However, this state originates from hybridization of metal states with the HATCN LUMO as we will demonstrate later in this work, and therefore we use the term "HIS" rather than "former LUMO" or "F-LUMO. On Ag(111), the HIS appears in the UPS spectrum as an additional intensity near $E_F$.[13,32] Likewise, when 1 ML HATCN is deposited on $n$-ML Ag/Cu(111), a similar increased intensity near $E_F$ is observed in the UP- (Figure 3(a), 3(b)) and ARPE-spectra (Figure 3(c), 3(d), Figure S2 and S3, Supplementary Information), which can again be reasonably assigned as the HIS. Note that the HIS is expected to be fundamentally similar in nature even when $n = 0$ (i.e. Cu(111)), since HATCN adsorption on Ag(111) and Cu(111) both show interfacial hybridization.[31] In our setup, the spectral shape of the HIS can be modeled as a Gaussian feature arising from the hybridization of the LUMO with the Ag states near $E_F$, cut off by the $E_F$.[32,34] Figures 3(c)-(d) show the ARPES of 10 ML Ag/Cu(111) and 1 ML HATCN/10 ML Ag/Cu(111). The features observed in the $n$-ML Ag/Cu(111) ARPES in the $\Gamma \to K$ direction and binding energy window of 0 to -2 eV are the SSS, the Ag $sp$-band and the Ag/Cu(111) quantum well states (QWS).[17,18] The latter appear only for Ag thicknesses ≥10 ML. As shown in Figure 3(d) and Figure S3(h), the HIS appears throughout the Brillouin zone in the ARPES spectra, but the intensities are the highest for $k_\parallel \geq 0.5$ Å$^{-1}$ in the $\Gamma \to K$ direction of the emerging Ag(111) Brillouin zone, an observation that can be attributed to the angle dependence of the photoemission intensity matrix of the HIS, the polarization and the angle of incidence of the He source, as has also been observed for other metal-organic interfaces.[35–37] The SSS disappears upon deposition of >1 ML HATCN(see Figure 3(d)). Additionally, for thicker Ag films ($n \geq 10$ ML), Umklapp (back-folded) bands of the Ag $sp$-band are also observed in the ARPES (Figure S2, Supplementary Information) which are formed due to the periodicity imposed by the HATCN superlattice.[38,39] From these UPS and ARPES data we find that the HIS remains, with binding energies between 0 and -0.5 eV for all thicknesses of the Ag film investigated. The formation of an ordered HATCN superstructure and the energetic proximity of the HIS to the $E_F$ are significant from the perspective of interfacial organic electronics, since both play a key role in determining charge transport through the organic molecule or layer.[1,9,40,41]

Having established the thickness-dependent surface electronic structure of the $n$-ML Ag/Cu(111) system and the existence of a HIS in the 1 ML HATCN/ $n$-ML Ag/Cu(111) system, we now investigate the thickness-dependence of the HIS on the underlying Ag film thickness. The UPS spectra ($I(E)$) of 1 ML HATCN/ $n$-ML

Ag/Cu(111) (see Figure 3(b)) were fitted using a simple model (equation ($S1$), see

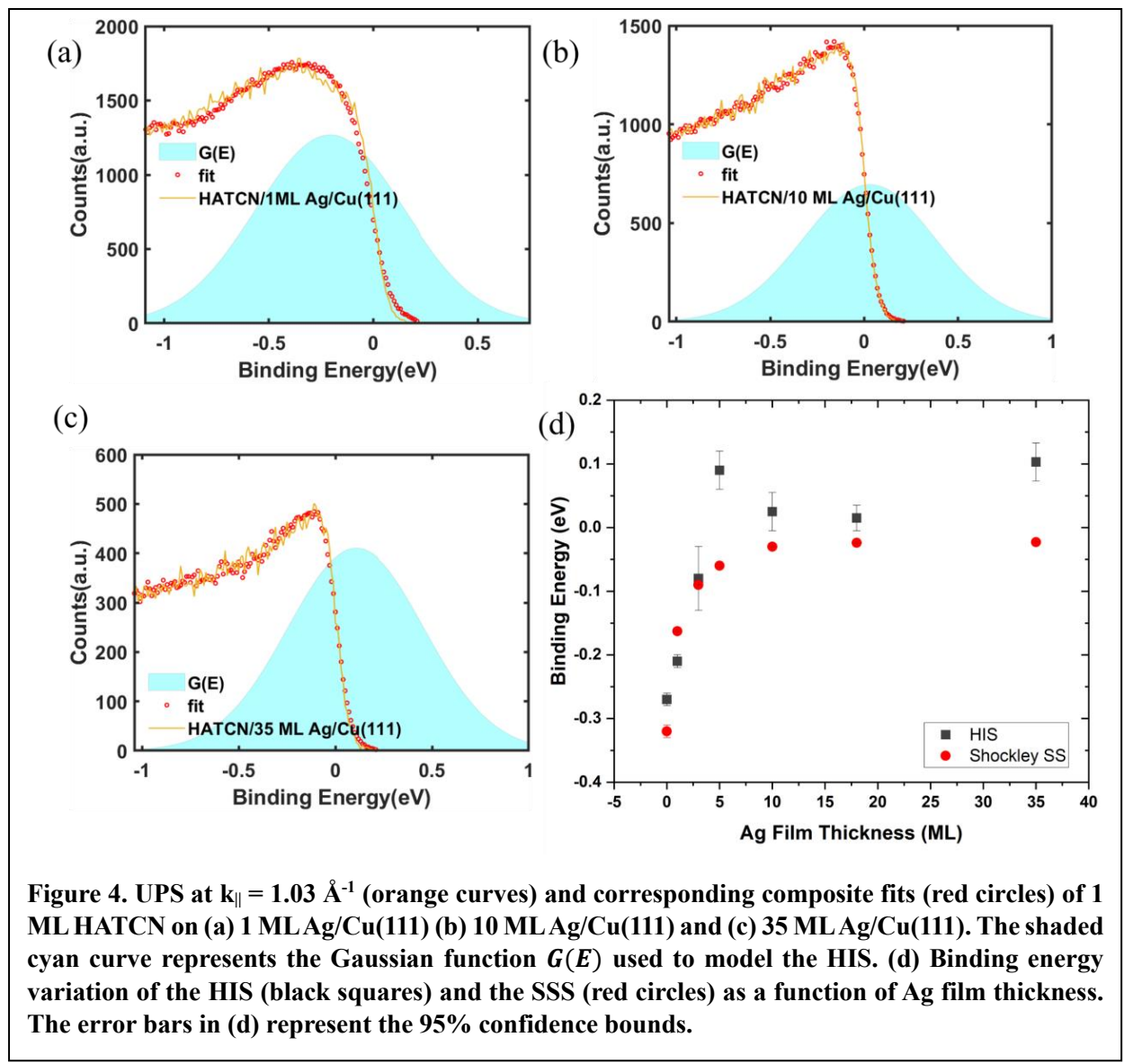

**Figure 4.** UPS at $k_\parallel$ = 1.03 Å$^{-1}$ (orange curves) and corresponding composite fits (red circles) of 1 ML HATCN on (a) 1 ML Ag/Cu(111) (b) 10 ML Ag/Cu(111) and (c) 35 ML Ag/Cu(111). The shaded cyan curve represents the Gaussian function $G(E)$ used to model the HIS. (d) Binding energy variation of the HIS (black squares) and the SSS (red circles) as a function of Ag film thickness. The error bars in (d) represent the 95% confidence bounds.

Supplementary Information for more details) that includes a Gaussian for the HIS, the Fermi-Dirac distribution, a background and the pristine *n*-ML Ag/Cu(111) spectrum. Figure 4(a)-(c) shows the fit results for *n* =1,10 and 35, where we only show the full fits (red circles) and the Gaussian representing the HIS (cyan shaded area). The fits are in excellent agreement with the experimental data (see also Figure S4, Supplementary Information). We use the Gaussian function $G(E)$ to model the HIS, and choose to assign the Gaussian center to the binding energy of the HIS. Figure 4(d) shows the systematic binding energy shift of the HIS as a function of Ag film thickness: The HIS has a binding energy of -0.26(1) eV on Cu(111) (0 ML Ag), decreases towards $E_F$ with increasing Ag film thickness, and finally reaches a value of 0.10(3) eV for 35 ML Ag/Cu(111). The HIS binding energy values are listed in Table S1 (Supplementary Information). The HIS does not show any dispersion within the error bounds of our fits (using equation (S1)) performed at different $k_\parallel$ values.

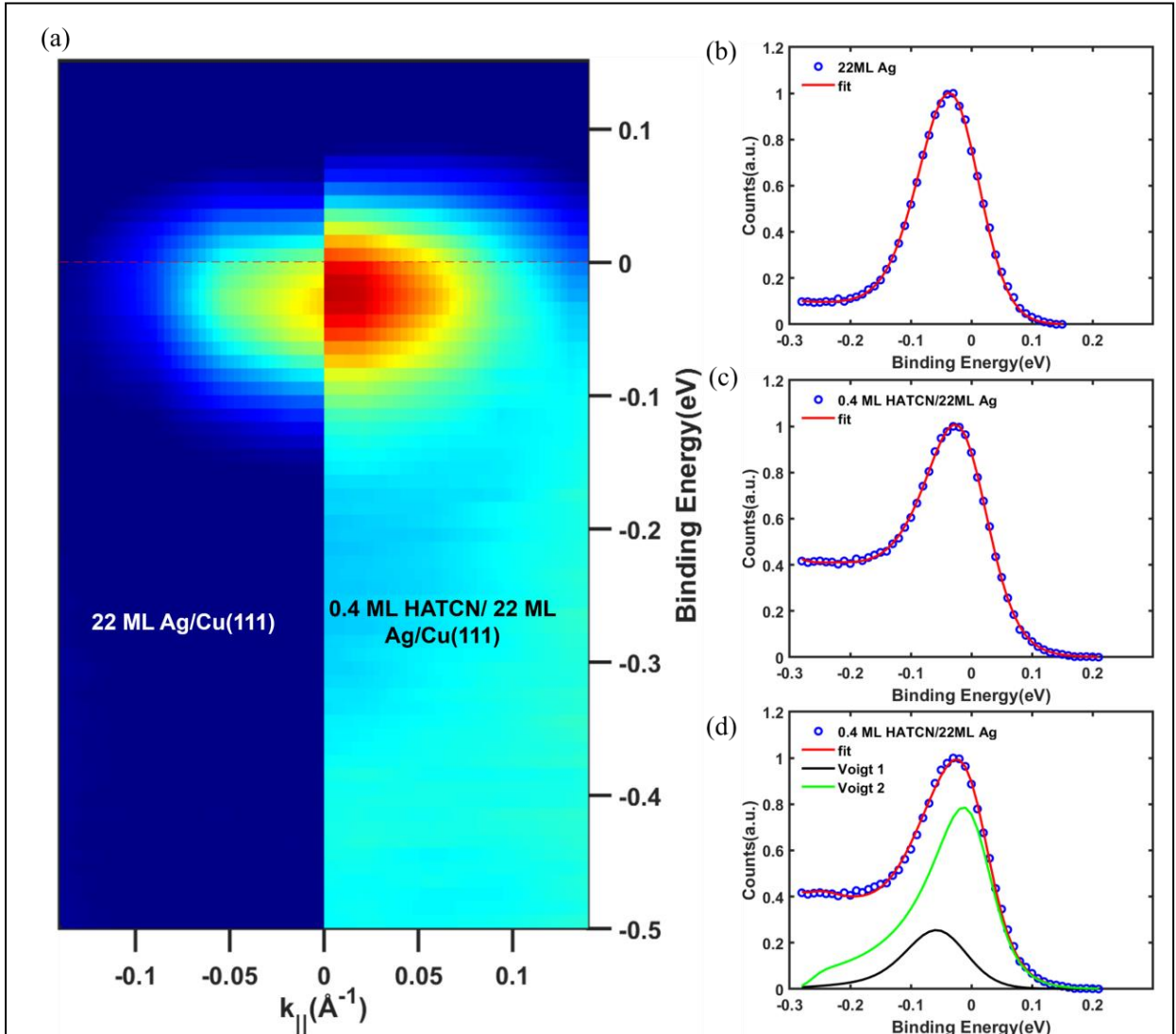

Figure 5. (a) Normalized ARPES of 22 ML Ag/Cu(111) (left panel) and 0.4 ML HATCN / 22 ML Ag/Cu(111) (right panel), showing the SSS evolution due to HATCN adsorption. (b,c) UPS at normal emission ($k_\parallel = 0$) (blue circles) and fits using a single Voigt function (red curves) of (b) 22 ML Ag/Cu(111) and (c) 0.4 ML HATCN/ 22 ML Ag/Cu(111). (d) Fit (red curve) using a sum of two Voigt functions (black (Voigt 1, for the bare Ag/Cu(111) SSS) and green (Voigt 2, for the shifted SSS) curves) of UPS at normal emission ($k_\parallel = 0$) of 0.4 ML HATCN/ 22 ML Ag/Cu(111) (blue circles). All fits also include background and the Fermi-Dirac distribution.

The results shown in Figure 4(d) and Table S1 reveal the following significant features of the 1 ML HATCN/ $n$-ML Ag/Cu(111) system: (i) The HIS binding energy can be tuned within a large energy window of ~360 meV, where it can lie well below $E_F$ (-0.26(1) eV, 0 ML Ag) or well above $E_F$ (0.10(3) eV, 35 ML Ag); and (ii) the thickness dependence of the HIS binding energy is not linear. Rather, it remarkably resembles the exponential behavior of the SSS binding energy with thickness. We propose that the origin of this behavior reflects the nature of such interface states in metal-organic interfacial systems: The existence of a charge-transfer derived HIS is widely reported in metal-organic interfaces,[11,13,34,35,42–48] and it was shown that the hybridization of the LUMO with the metal states near $E_F$ is central to such HIS formation.[35,43,47,49] Since the thickness and structure of the HATCN layer was kept constant in our study (as verified by LEED, see Figure S5 in the Supplementary Information), we may safely rule out a variation in the intermolecular hybridization as the origin of this effect. Energy level alignment of HATCN/Ag(111)[12,13] shows that the HATCN LUMO sits right at $E_F$. Therefore, we propose that a variation in the DOS near $E_F$ of the substrate states participating in the hybridization with the LUMO is the origin of the thickness dependent binding energy shift of the HIS. We suggest that the SSS of the $n$-ML Ag/Cu(111) system hybridizes with the HATCN LUMO to form a thickness dependent HIS, and we justify this hypothesis next. Hybridization of SSS with molecular occupied/unoccupied states in hybrid interfaces has been studied both experimentally and theoretically.[10,50–55] For PTCDA/Ag(111), a new interfacial state was observed with an SSS-like wavefunction inside Ag(111) and molecular states outside.[50,51] In O/Pt(111), preferential hybridization between O states and Pt(111) surface states was observed,[52] and in fact the Ag/Cu(111) thin film system itself has been used to control the self-assembly of adsorbed 2H-TPP molecules by making use of the thickness dependence of the SSS Fermi wavelength, though the electronic structure was not investigated.[10] Beyond the appearance of a HIS, a possible signature of SSS-molecule hybridization in such systems is a shift of the SSS closer to $E_F$.[50,52] Therefore we studied the effect of HATCN adsorption on the SSS using UPS and ARPES, focusing on sub-ML HATCN films (0.4 ML HATCN) since the SSS shifts above $E_F$ for 1 ML HATCN.

Figure 5(a), right panel, shows the evolution of the SSS for this scenario, for the case of 22 ML Ag/Cu(111). In these normalized spectra, we observe a shift to lower binding energy and consequently an increase in the SSS DOS near and above $E_F$ compared to bare 22 ML Ag/Cu(111) (Figure 5(a), left panel), without a significant

shift of the *overall* spectra (as expected for a full ML HATCN coverage, see Figure 3(d) and Figure S2). To model this situation in more detail, we considered two contrasting physical scenarios to understand the ARPES observations: (A) No hybridization occurs between the SSS and molecular states, and the SSS only gets broadened due to overlayer scattering without a binding energy shift. (B) Charge-transfer and hybridization between the SSS and molecules leads to a shift of the SSS binding energy closer to $E_F$. The bare 22 ML Ag/Cu(111) and the 0.4 ML HATCN/ 22 ML Ag/Cu(111) SSS spectra for scenario (A) is then modeled using a single Voigt function whose Gaussian width we expect to increase upon HATCN deposition, whereas we use a sum of two Voigt functions for scenario (B) to account for SSS photoemission from bare Ag/Cu(111) and a shifted SSS from HATCN/Ag/Cu(111) patches, since the HATCN coverage is below 1 ML. Please note that the shifted SSS and the HIS are two different states. Equation ($S3$) (Supplementary Information) describes the model used to fit our normal emission UPS data. The bare 22 ML Ag/Cu(111) SSS fit yields a Voigt center $C = -0.037(2)$ eV and half-width $W = 0.041(5)$ eV (Figure 5(b)). The 0.4 ML HATCN/ 22 ML Ag/Cu(111) SSS fit (Figure 5(c)) for scenario (A) yields $C = -0.011(2)$ eV and $W = 0.042(4)$ eV. The increase in the half-width (1 meV) due to HATCN adsorption is within the error bar while the shift in the SSS binding energy towards $E_F$ is significant (26 meV), indicating that scenario (A) is unlikely. For scenario (B), we fit the same 0.4 ML HATCN/ 22 ML Ag/Cu(111) SSS using equation ($S3$) with $m = 2$, where the center ($C_1$) for one of the Voigt functions (Voigt 1) was fixed at -0.037 eV to model emission from bare 22 ML Ag/Cu(111) patches. Our model fits the data well (Figure 5(d) and Table 2) and yields $C_2 = -0.009(2)$ eV and half-width $W_2 = 0.034(7)$ eV for the second Voigt function (Voigt 2). This suggests that the 0.4 ML HATCN/ 22 ML Ag/Cu(111) SSS spectrum is a sum of bare Ag/Cu(111) SSS emission and the emission from a shifted SSS interacting with the HATCN molecules. Our results therefore indicate that scenario (B) is a more likely explanation of the observed evolution of the HATCN/Ag/Cu(111) interface, indicating that the SSS shifts towards $E_F$ by 28 meV.

**Table 2. SSS Fit Results of 22 ML Ag/Cu(111) and 0.4 ML HATCN/ 22 ML Ag/Cu(111)**

| System | Fitted Voigt Parameters (Voigt Center ($C$) and HWHM ($W$)) for $m = 1$ | Fitted Voigt Parameters (Voigt Center ($C_1,C_2$) and HWHM ($W_1,W_2$)) for $m = 2$ |
|---|---|---|
| 22 ML Ag/Cu(111) | $C$ = -0.037(2) eV  $W$ = 0.041(5) eV | - |
| 0.4 ML HATCN/ 22 ML Ag/Cu(111) | $C$ = -0.011(2) eV  $W$ = 0.042(4) eV | $C_1$ = -0.037 eV (fixed)  $W_1$ = 0.039(17) eV  $C_2$ = -0.009(2) eV  $W_2$ = 0.034(7) eV |

To summarize, upon adsorption the HATCN LUMO hybridizes with the $n$-ML Ag/Cu(111) SSS, undergoes charge-transfer and forms a HIS. This is accompanied by shift of the $n$-ML Ag/Cu(111) SSS towards $E_F$ (e.g. by 28 meV, for $n$ =22 ML). The binding energy of the SSS is tunable between -0.32(1) eV and -0.023(1) eV for the system studied, and we find that the HIS binding energy varies between -0.26(1) eV and 0.10(3) eV with increasing Ag film thickness, in a functional form that mimics that of the HIS. Underpinning this thickness dependence of the HIS is the significant contribution of the SSS in HIS formation, as suggested in previous works.[10,50–53] The significance of our results therefore is that we are able to directly test this suggestion and that we are consequently able to demonstrate a way to precisely adjust the DOS of the metal substrate near $E_F$ and use the hybrid nature of the HIS to create tunable interfacial states in this metal-organic thin film systems.

Other possibilities such as hybridization with the Ag/Cu(111) QWS or with the $sp$-band can be ruled out since (1) the thickness-dependent HIS binding energy shift is already observed for $n$ <10 ML Ag films, well below the minimum thickness at which the $v = 1$ QWS is first observed in UPS,[18] and (2) the Ag $sp$-band dispersion in $n$-ML Ag/Cu(111) is expected to be the same as bulk Ag(111), in agreement with previous work[56] and our own ARPES and UPS results (see Figure S1,

Supplementary Information). Hence neither can explain the thickness-dependent binding energy shift of the HIS. The fact that the binding energy shift is only observed for the HIS and not for other states such as the HATCN HOMO (binding energy ~ -4.2 eV) further supports our central hypothesis. Since the work function of the $n$-ML Ag/Cu(111) is always 4.50(5) eV, we rule out surface potential variations as a possible source of the HIS binding energy shift.

We also considered the role of the Ag/Cu(111) surface structure on the HATCN adsorption and HIS formation. It is well known that the 1 ML Ag/Cu(111) surface shows a (9 × 9) reconstruction due to the lattice mismatch between Ag and Cu (~13%).[21,22] When the system is prepared at 298 K (as is the case in the present work), the surface shows triangular corrugations due to the formation of dislocation loops in the first Cu(111) layer. Consequently, states that are confined near the surface may be influenced by this surface modification. Indeed, previous studies have reported the development of a SSS band-gap and Fermi surface modification in the 1 ML Ag/Cu(111) system due to back-folding effects.[22] Preferential adsorption of single molecular adsorbates on the hcp sites on this surface has also been observed.[24] It is possible that the HIS wavefunction in the 1 ML HATCN/ 1 ML Ag/Cu(111) system extends into the Cu(111) layer, which may explain the higher binding energy of the HIS relative to the SSS binding energy observed only for the 1 ML Ag case. However, we do not observe any significant modification of adsorption properties for the monolayer HATCN on the reconstructed 1 ML Ag surface compared to thicker Ag films as tested by LEED. All diffraction patterns of the HATCN lattice on all Ag thicknesses measured are identical within the error bounds of our apparatus (see Figure S5, Supplementary Information). This suggests that the macroscopic electronic and structural properties of the self-assembled HATCN layer are robust to the surface modifications in Ag/Cu(111) even down to a single Ag monolayer, which could be significant from the perspective of creating molecule-metal interfacial heterostructures where the overlayer structure is consistent, long-range ordered and independent of underlying layer thicknesses.

In conclusion, our results show the formation of a HIS upon adsorption of HATCN on thin films (1-35 ML) of Ag on Cu(111). We show that this HIS is tunable as a result of hybridization and charge-transfer between the Ag/Cu(111) SSS and the HATCN LUMO. The thickness dependence of the SSS binding energy allows us to finely control the DOS near $E_F$, and in turn to vary the binding energy of the HIS over a large energy range (~360 meV) near $E_F$. The formation of a well-ordered organic superstructure and a highly tunable charge-transfer state is highly significant and can be generalized to a multitude of thin film metal-organic interfaces for controlling interfacial energy level alignment and charge-transfer in organic electronics.

Supporting Information

Materials, Characterization, ARPES, UPS, Fitting, LEED

Acknowledgments

This research was supported by the National Science Foundation under Grant No. NSF CHE-1954571 and by the Galileo Circle Scholarship from the College of Science at the University of Arizona. We thank Prof. Dr. Torsten Fritz for providing us with access to LEEDCal and LEEDLab for LEED structure analysis.

Supplementary Material

Tailoring Charge-Transfer at Metal-Organic Interfaces Using Designer Shockley Surface States

Anubhab Chakraborty,[1] Oliver L.A. Monti[1,2]*

[1]Department of Chemistry and Biochemistry, University of Arizona, Tucson, Arizona 85721, United States

[2]Department of Physics, University of Arizona, Tucson, Arizona 85721, United States

## S1: Band Structure of *n*-ML Ag/Cu(111)

Figure S1 shows our ARPES results for *n*-ML Ag/Cu(111). The thickness dependent

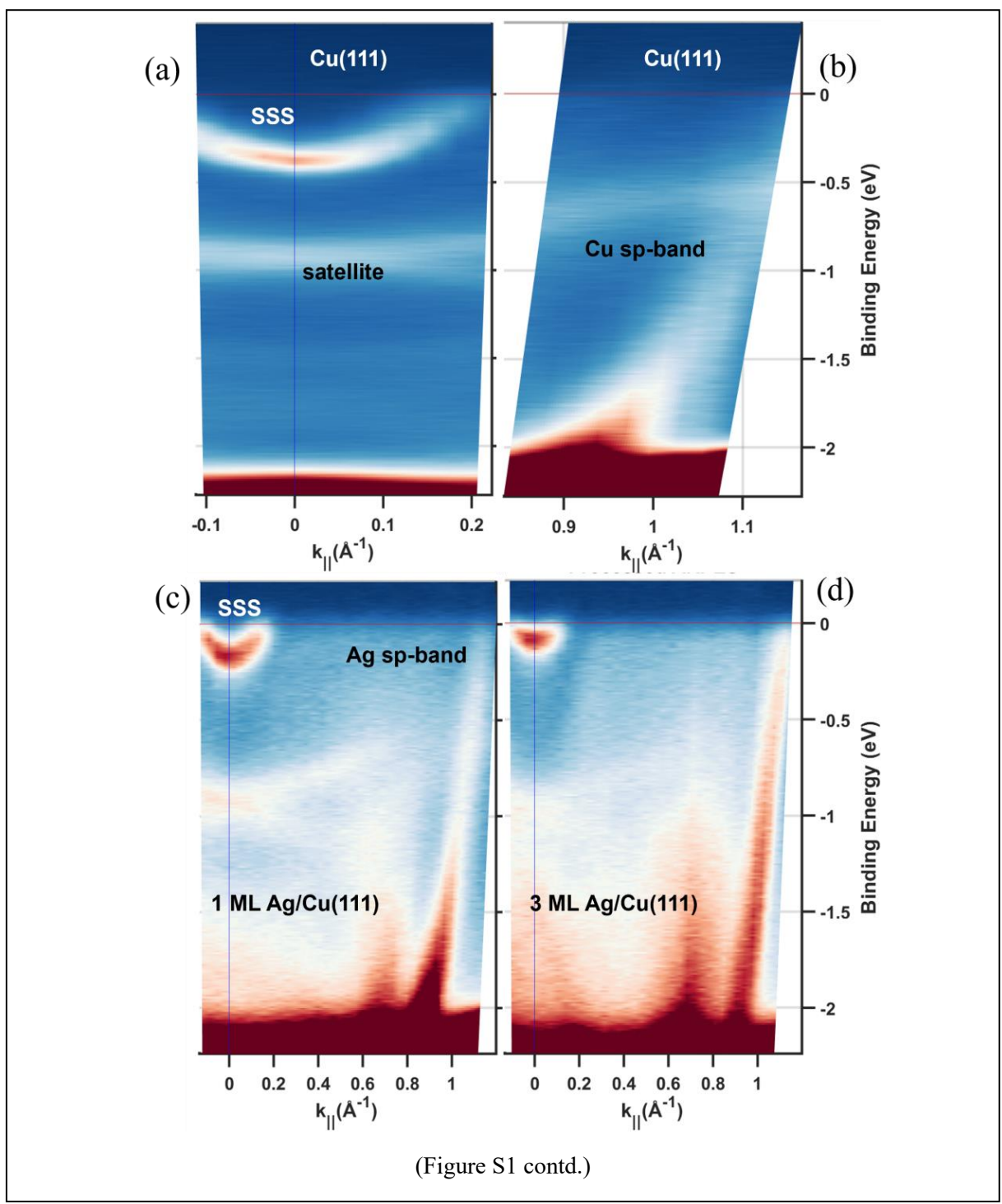

(Figure S1 contd.)

variation of the SSS is clearly observed. Further, Ag QWS are observed for Ag film thicknesses ≥10 ML. The *sp*-bands of Cu (Figure S1(b)) and Ag (Figure S1(c)-(h)) are also observed. While the intensity increases with increasing film thickness, there is no variation in the binding energy or band dispersion of the *sp*-band with Ag film thickness. Some weaker satellites of the Cu *d*-bands are also observed (Figure S1(a), (c)-(e)), arising from He Iβ emission.

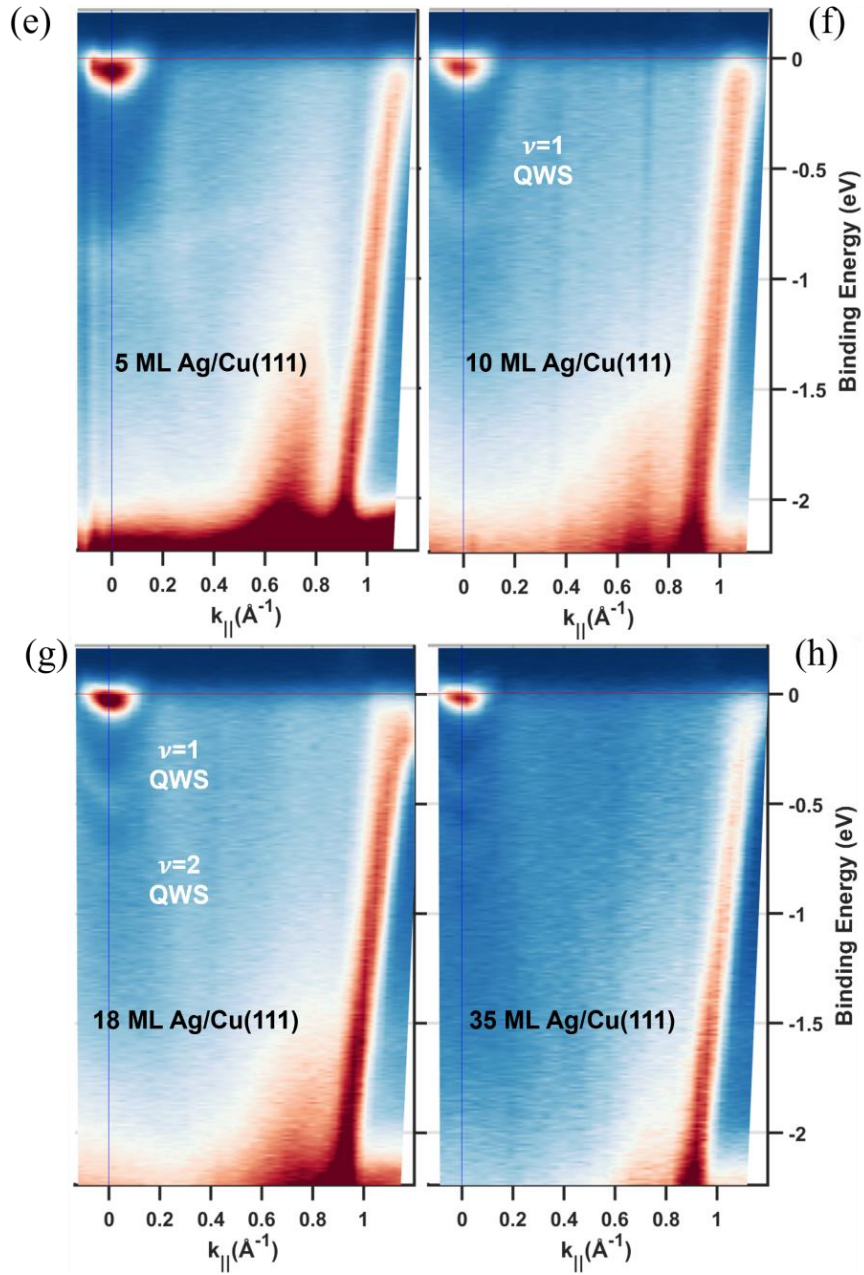

**Figure S1.** ARPES maps in the $\Gamma \rightarrow K$ direction of (a) Cu(111) (near $\Gamma$) (b) Cu(111) (higher $k_\parallel$) (c) 1 ML Ag/Cu(111) (d) 3 ML Ag/Cu(111) (e) 5 ML Ag/Cu(111) (f) 10 ML Ag/Cu(111) (g) 18 ML Ag/Cu(111) and (h) 35 ML Ag/Cu(111). The prominent features in this binding energy range are the Shockley Surface State (SSS), *sp*-band, Quantum Well States (QWS) and satellite features arising from He I$\beta$ emission.

S2: Band Structure of 1 ML HATCN/ *n*-ML Ag/Cu(111)

Figure S2 shows our ARPES results for 1 ML HATCN/*n*-ML Ag/Cu(111). The SSS is no longer observed upon growth of 1 ML HATCN. The HIS appears as an increased intensity at higher momenta ($k_\parallel \geq 0.5$ Å$^{-1}$). The HIS intensity is weaker on

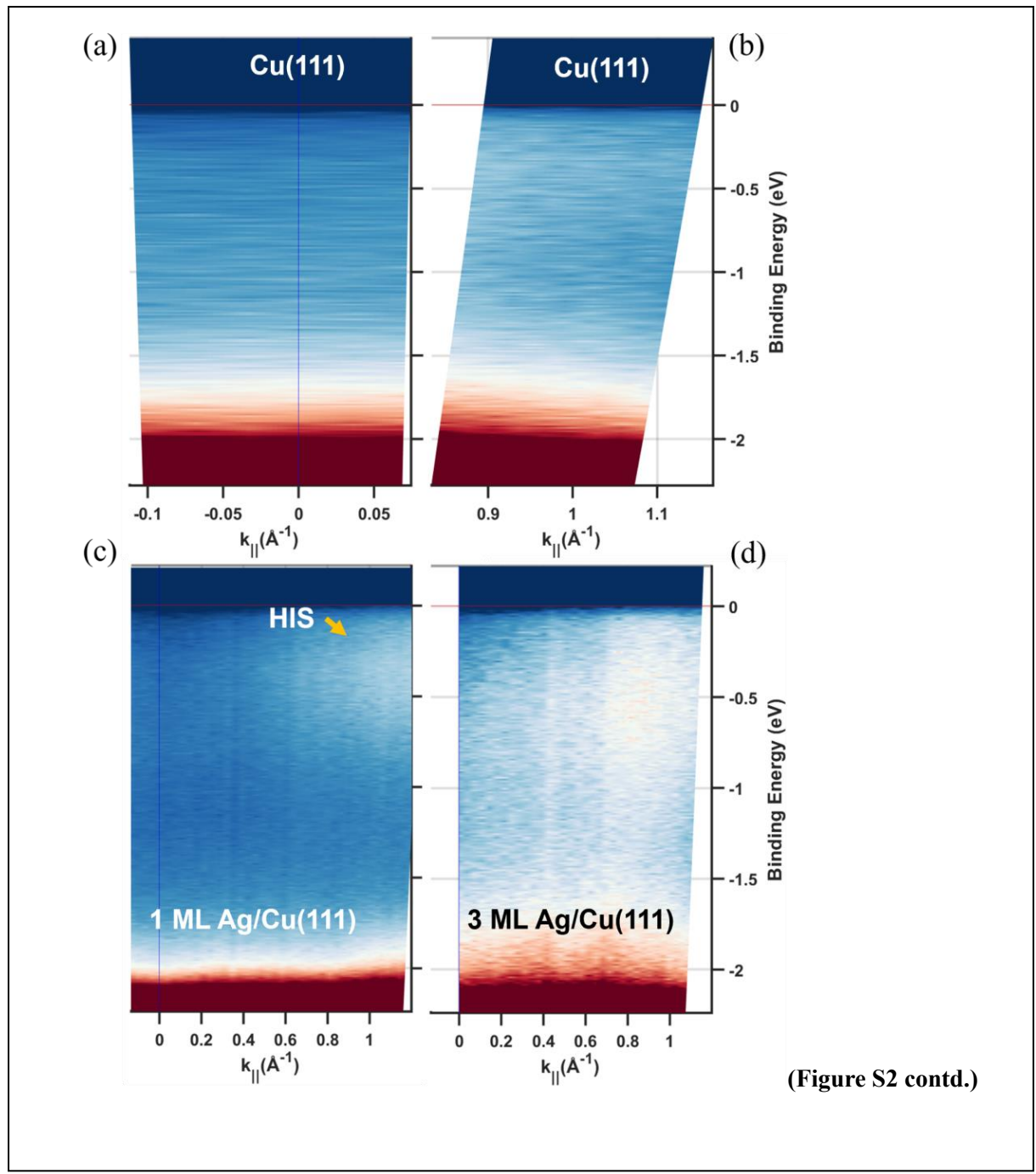

(Figure S2 contd.)

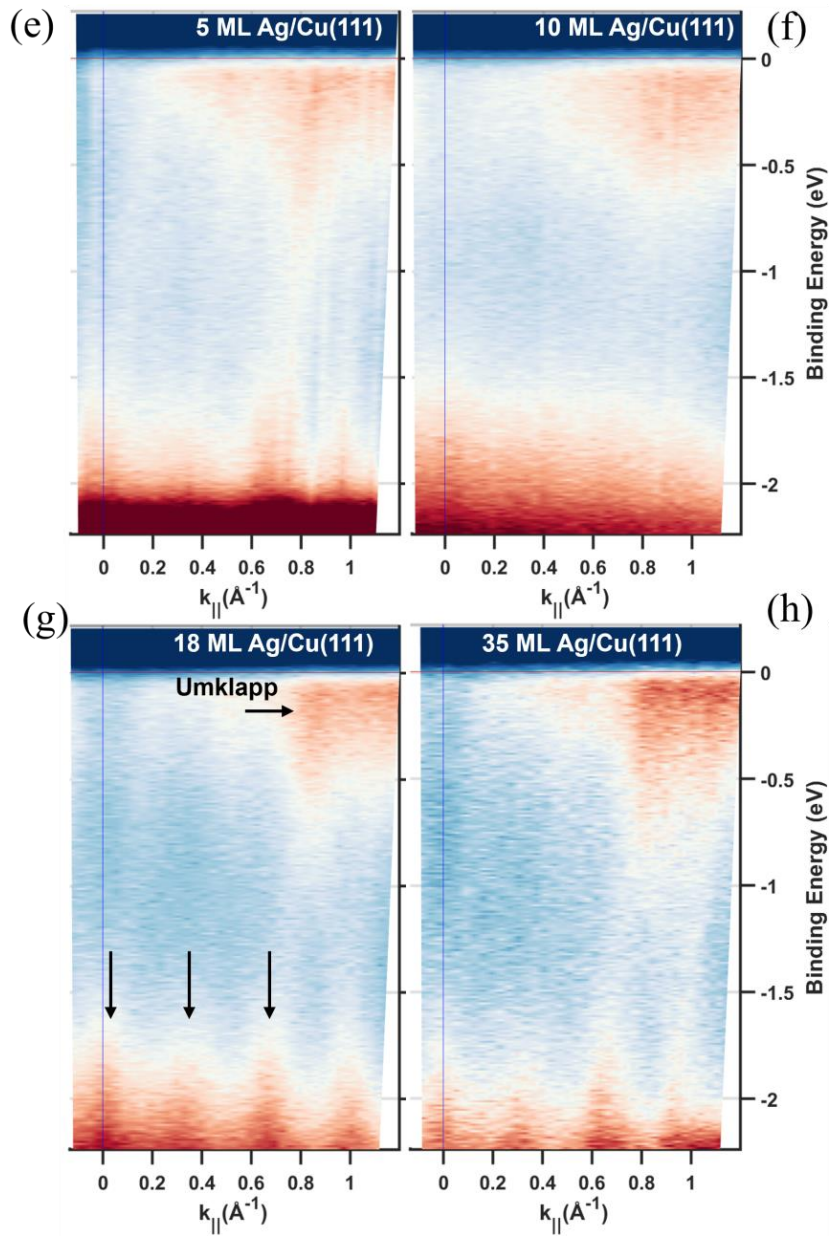

**Figure S2.** ARPES maps in the $\Gamma \rightarrow K$ direction of 1 ML HATCN on (a) Cu(111) (near $\Gamma$) (b) Cu(111) (higher $k_\parallel$) (c) 1 ML Ag/Cu(111) (d) 3 ML Ag/Cu(111) (e) 5 ML Ag/Cu(111) (f) 10 ML Ag/Cu(111) (g) 18 ML Ag/Cu(111) and (h) 35 ML Ag/Cu(111). The prominent features in this binding energy range are the Hybrid Interface State (HIS) and the Umklapp sp-bands (indicated by arrows) observed for thicker Ag films (g)-(h). The slight slope of $E_F$ with $k_\parallel$ is attributed to a small tilt of the Cu crystal on the sample holder.

Cu(111) (Figure S2(b)) compared to Ag/Cu(111) (Figure S2(c)-(h)), and is more clearly depicted in the UPS plots (Figure S3(a)). Umklapp (or back-folded) bands of the Ag *sp*-band are also observed for thicker Ag films (Figure S2(g)-(h), blue arrows), formed due to the periodicity imposed by the HATCN superlattice.

## S3: UPS of 1 ML HATCN/*n*-ML Ag/Cu(111)

Figure S3 shows the appearance of a HIS in the UPS after deposition of 1 ML HATCN on *n*-ML Ag/Cu(111) (see main text for spectral shape analysis). From the bare $n$-ML Ag/Cu(111) spectra (blue curves, Figure S3), we also determine that there is no systematic shift of the DOS for the Ag $sp$-band with increasing Ag film thickness. Due to the different band structures of Ag/Cu(111) and bare Cu(111), slightly different $k_{||}$ values (1.03 Å$^{-1}$ and 0.90 Å$^{-1}$ respectively) were selected for analysis of the HIS. The HIS shows an angle-dependent photoemission intensity, having higher intensities at higher $k_{||}$ values (Figure S3(h)).

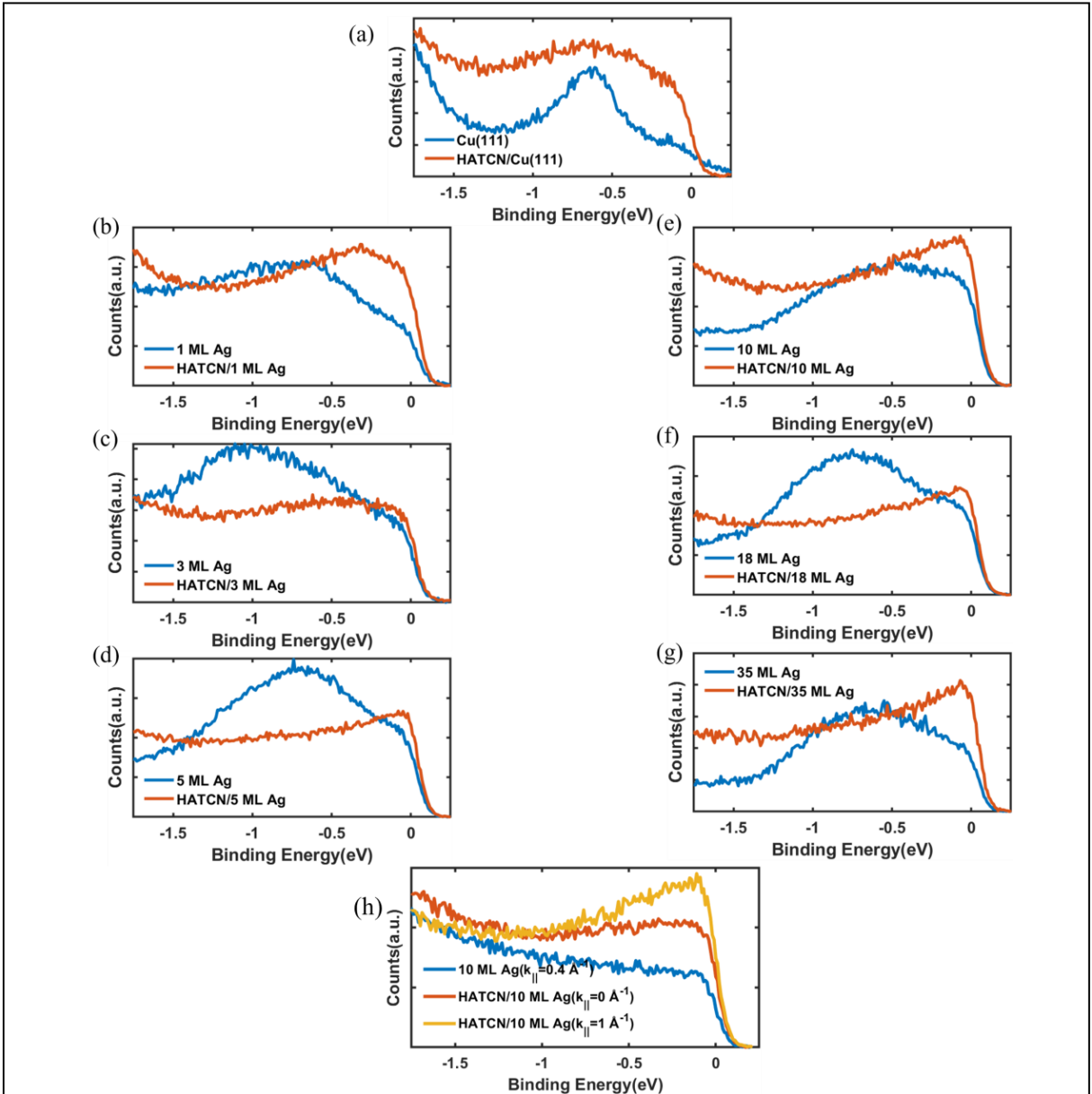

**Figure S3.** UPS at $k_{\parallel}$ = 0.90 Å$^{-1}$ of (a) Cu(111) (blue curve) and 1 ML HATCN on Cu(111) (red curve). UPS at $k_{\parallel}$ = 1.03 Å$^{-1}$ of *n*-ML Ag/Cu(111) (blue curves) and 1 ML HATCN on *n*-ML Ag/Cu(111) (red curves) for *n* = 1- 35 (b)-(g). The *n*-ML Ag/Cu(111) spectra show the *sp*-band DOS for (a) Cu and (b)-(g) Ag, and the HATCN/*n*-ML Ag/Cu(111) spectra shows the extra DOS due to the Hybrid Interface State (HIS). (h) UPS plots of HATCN/ 10 ML Ag/Cu(111) showing the HIS intensity variation at $k_{\parallel}$ = 0 Å$^{-1}$ (red) and $k_{\parallel}$ = 1 Å$^{-1}$ (yellow). The 10 ML Ag/Cu(111) plot at $k_{\parallel}$ = 0.4 Å$^{-1}$ (blue) shows the substrate Fermi edge.

S4: UPS fits of 1 ML HATCN/ $n$-ML Ag/Cu(111)

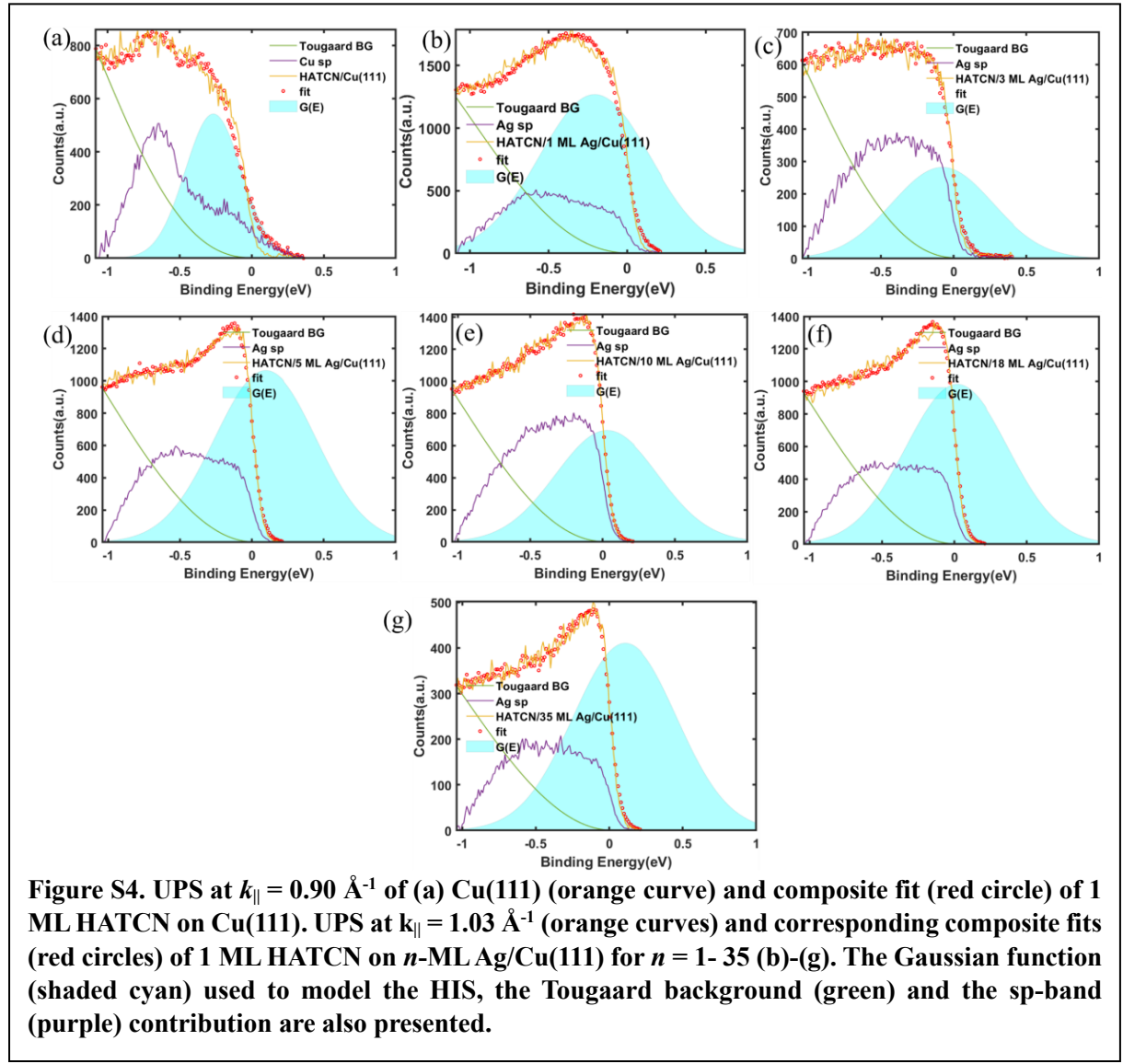

**Figure S4.** UPS at $k_{||}$ = 0.90 Å$^{-1}$ of (a) Cu(111) (orange curve) and composite fit (red circle) of 1 ML HATCN on Cu(111). UPS at $k_{||}$ = 1.03 Å$^{-1}$ (orange curves) and corresponding composite fits (red circles) of 1 ML HATCN on $n$-ML Ag/Cu(111) for $n$ = 1- 35 (b)-(g). The Gaussian function (shaded cyan) used to model the HIS, the Tougaard background (green) and the sp-band (purple) contribution are also presented.

Figure S4 shows the UPS fits of 1 ML HATCN/$n$-ML Ag/Cu(111) to extract the HIS binding energies. We chose the spectra at $k_{||}$ = 0.90 Å$^{-1}$ for bare Cu(111) and at $k_{||}$ =

1.03 Å⁻¹ for *n*-ML Ag/Cu(111) to avoid additional DOS in the spectra due to the Umklapp *sp*-band features. The small difference in $k_{\parallel}$ does not affect the fits since the HIS is non-dispersive. The following model was used for fitting the UPS data ($I(E)$):

$$I(E) = G(E) \times FD(E) + c\left(I_{Ag}(E) - BG_{Ag}(E)\right) + BG_{HATCN}(E) \qquad (S1)$$

where $G(E)$ is a Gaussian distribution, $FD(E)$ is the Fermi-Dirac distribution to account for the spectral shape near $E_F$, $I_{Ag}(E)$ is the pristine *n*-ML Ag/Cu(111) spectrum which takes into account the *sp*-band contribution, $BG_{Ag}(E)$ and $BG_{HATCN}(E)$ are the Tougaard background[1] of the *n*-ML Ag/Cu(111) and the 1 ML HATCN/ *n*-ML Ag/Cu(111) respectively, and $c$ is a scaling parameter for the fit. We note in passing that the results do not depend significantly on the background model used and similar results are obtained e.g. for a Shirley background. Although the half-width at half maximum (HWHM) of the Gaussian was $G(E)$ was allowed to vary between 0-0.5 eV (typical values for organic adsorbate features in UPS) to improve fit quality, our results in Table S1 indicate that for most thicknesses, the HWHM is fixed at the upper bound of 0.5 eV. The Fermi-Dirac distribution takes the form:

$$FD(E) = \left[1 + e^{\frac{E}{k_B T_{eff}}}\right]^{-1} \qquad (S2)$$

where $k_B$ is the Boltzmann constant and $T_{eff}$ captures the instrument broadening, which was estimated by fitting the Fermi edge of *n*-ML Ag/Cu(111) UPS at $k_{\parallel} = 0.5$ Å⁻¹. For our setup, $k_B T_{eff} \sim 40$ to $55$ meV. Our model fits the data well in general, and the binding energies and HWHM of the HIS are listed in Table S1 below. From Table S1 and Figure 4(d) in the main text, we conclude that the estimated HIS binding energy for the 1 ML HATCN/5 ML Ag/Cu(111) case might be an outlier (slightly too high), although the reason for that is presently unclear.

**Table S1. HIS Binding Energy and HWHM as a function of the Ag film thickness**

| Ag film thickness (ML) | HIS Binding Energy (eV) | HWHM (eV) |
|---|---|---|
| 0 | -0.26(1) | 0.27(1) |
| 1 | -0.21(1) | 0.5 (fixed at bound) |
| 3 | -0.08(5) | 0.49(5) |
| 5 | 0.09(3) | 0.5 (fixed at bound) |
| 10 | 0.03(3) | 0.5 (fixed at bound) |
| 18 | 0.02(2) | 0.5 (fixed at bound) |
| 35 | 0.10(3) | 0.5 (fixed at bound) |

## S5: UPS fits of 0.4 ML HATCN/ 22 ML Ag/Cu(111)

The following model was used to fit our normal emission UPS data (Figure 1(c) and Figures 5(b)-(d), main text):

$$I(E) = \left(\sum_{i=1}^{m} V_i(E)\right) \times FD(E) + BG(E), \quad m = 1 \text{ or } 2 \quad (S3)$$

where $V_i(E)$ is a Voigt profile, $FD(E)$ is the Fermi-Dirac distribution described in equation (2), $BG(E)$ is the Tougaard background and $m$ =1 or 2 for scenario (A) or (B) respectively, as described in the main text. The center for the Gaussian and the Lorentzian profiles in $V_i(E)$ was restricted to the same value to simulate Gaussian broadening of a Lorentzian SSS line profile. The Voigt width was calculated using the method described in previous works,[2] and is listed in Table 3. For scenario (A) ($m = 1$), all other fit parameters were unrestricted. For scenario (B) ($m = 2$), the Voigt center $C_1$ and the HWHM $W_1$ were restricted to -0.037(2) eV and 0.036(6) eV to model emission from bare 22 ML Ag/Cu(111) patches. Our model fits the data well, and the results and conclusions are shown in the main text (Figure 5(b)-(d), Table 3). This model ($m = 1$) was also used to find the *n*-ML Ag/Cu(111) SSS binding energies, as shown in Figure 1(b), (c) and Table 1, main text.

S5: LEED images of 1 ML HATCN/ *n*-ML Ag/Cu(111)

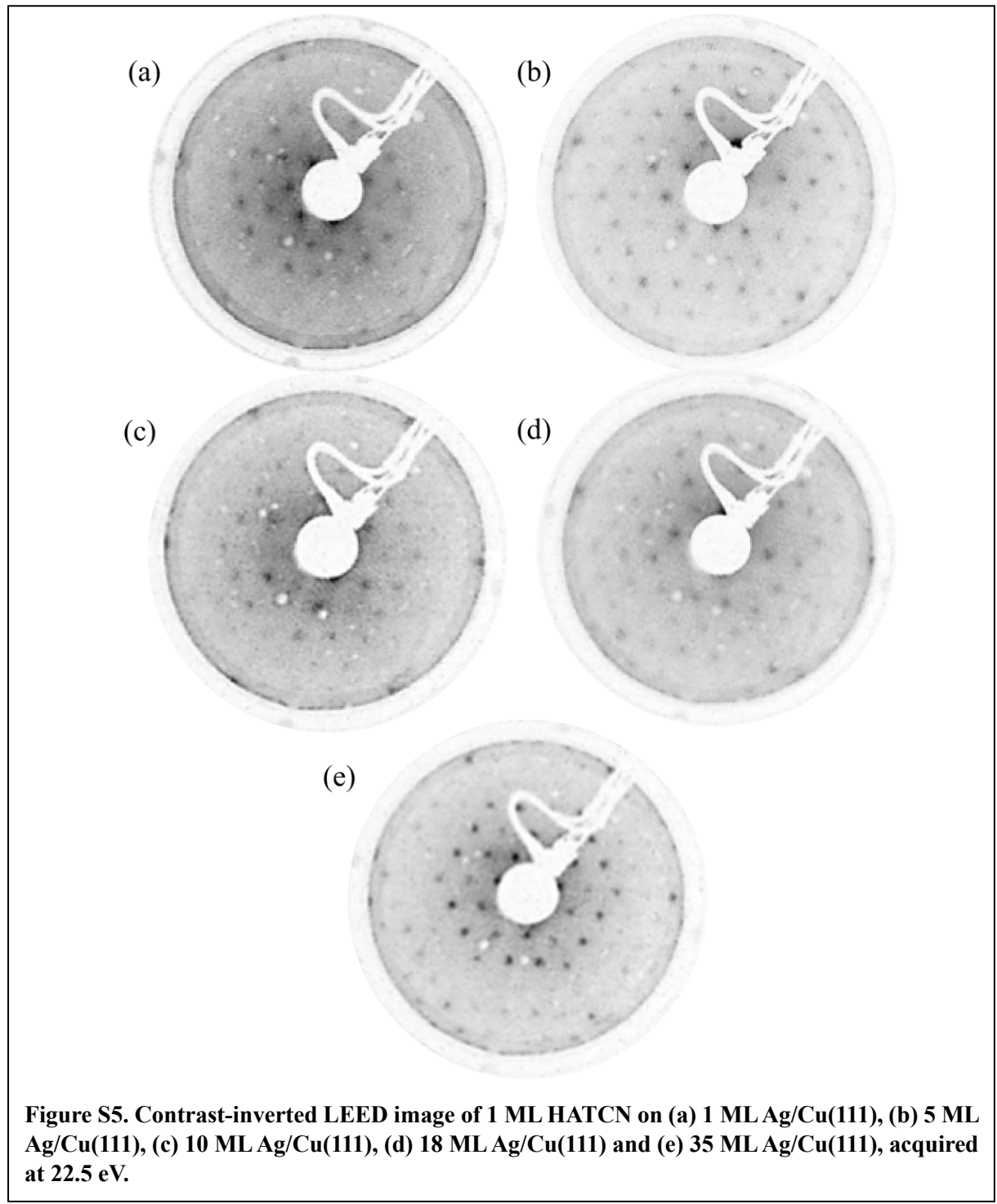

**Figure S5. Contrast-inverted LEED image of 1 ML HATCN on (a) 1 ML Ag/Cu(111), (b) 5 ML Ag/Cu(111), (c) 10 ML Ag/Cu(111), (d) 18 ML Ag/Cu(111) and (e) 35 ML Ag/Cu(111), acquired at 22.5 eV.**

Figure S5 shows the LEED images of 1 ML HATCN on 1 ML, 5 ML, 10 ML, 18 ML and 35 ML Ag/Cu(111). In all cases, we clearly observe the (7 × 7) HATCN superstructure. The unit cell structures for the Ag and HATCN unit cells are shown in Figure 2 in the main text. These images indicate the following important features of HATCN adsorption on Ag/Cu(111): (1) HATCN forms an ordered superstructure on this surface, (2) the adsorption structure of HATCN on Ag(111) and Ag thin films on Cu(111) are identical, and (3) the (9 × 9) reconstruction on the 1 ML Ag/Cu(111) surface does not significantly affect the HATCN superstructure, since the (7 × 7) LEED pattern is also observed for 1 ML HATCN on 1 ML Ag/Cu(111).